\documentclass{ifacconf}

\pdfoutput=1

\usepackage{natbib}        % required for bibliography
\usepackage[pdftex]{graphicx}
\usepackage{amsmath,amsfonts,amssymb}
\usepackage{float}
\usepackage[dvipsnames]{xcolor}
\usepackage{booktabs}
\usepackage[shortlabels]{enumitem}
\usepackage{subcaption}
\usepackage[left=43pt,right=43pt,top=106.75pt,bottom=39pt]{geometry} % Fix margins

\newtheorem{theorem}{Theorem}

\newtheorem{remark}{Remark}

\graphicspath{{./Figures/}}			% Path of figures

\newcommand{\mc}[1]{\mathcal{#1}}
\newcommand{\mbb}[1]{\mathbb{#1}}
\newcommand{\mrm}[1]{\mathrm{#1}}
\newcommand{\RHinf}{\mc{RH}_{\infty}}
\newcommand{\bbR}{\mbb{R}}

\newcommand{\Set}[2]{\ensuremath{\left\{ \vphantom{#2} #1 \right. \, \left| \, \vphantom{#1} #2 \right\}}}

\makeatletter
\g@addto@macro\normalsize{%
  \setlength\abovedisplayskip{.3em}
  \setlength\belowdisplayskip{.3em}
  \setlength\abovedisplayshortskip{.3em}%
  \setlength\belowdisplayshortskip{.3em}%
  \setlength\parskip{.3em}
}

\begin{document}
\begin{frontmatter}

%\title{Data-Driven Linear Parameter-Varying Control of a Control Moment Gyroscope\thanksref{footnoteinfo}} 
\title{Frequency-Domain Data-Driven Controller Synthesis for Unstable LPV Systems\thanksref{footnoteinfo}} 

\thanks[footnoteinfo]{This work has received funding from the European Research Council
(ERC) under the European Union's Horizon 2020 research and innovation
programme (grant agreement nr. 714663).}

\author[First]{Tom Bloemers} 
\author[First,Third]{Roland T\'oth} 
\author[Second]{Tom Oomen}

\address[First]{Control Systems Group, Department of Electrical Engineering, Eindhoven University of Technology, 5612 AE Eindhoven, The Netherlands (e-mail: \{t.a.h.bloemers, r.toth\}@tue.nl)
}
%\address[Second]{Control Systems Group, Department of Electrical Engineering, Eindhoven University of Technology, 5612 AE Eindhoven, The Netherlands.}
\address[Second]{Control Systems Technology, Department of Mechanical Engineering, Eindhoven University of Technology, 5612 AE Eindhoven, The Netherlands (e-mail: t.a.e.oomen@tue.nl).}

\address[Third]{Systems and Control Laboratory, Institute for Computer Science and Control, Kende u. 13-17, H-1111 Budapest, Hungary.}

\begin{abstract}                % Abstract of not more than 250 words.
Synthesizing controllers directly from frequency-domain measurement data is a powerful tool in the linear time-invariant framework. Ever-increasing performance requirements necessitate extending these approaches to account for plant variations. The aim of this paper is to develop frequency-domain analysis and synthesis conditions for local internal stability and $\mc{H}_\infty$-performance of single-input single-output linear parameter-varying systems. The developed synthesis procedure only requires frequency-domain measurement data of the system and does not need a parametric model of the plant. The capabilities of the synthesis procedure are demonstrated on an unstable nonlinear system.
\end{abstract}

%\begin{keyword}
%\end{keyword}

\end{frontmatter}
%===============================================================================

%===============================================================================
\section{Introduction}
\label{sec:introduction}
Frequency response function (FRF) measurements have traditionally been used to manually design controllers directly from measurement data. A frequency response function estimate provides an accurate nonparametric description of the system that is relatively fast and inexpensive to obtain \citep{System_Identification_Frequency_Domain}. This has enabled the use of classical techniques such as loop-shaping, alongside graphical tools including the Bode diagram or Nyquist plot, to design such controllers \citep{maciejowski1989multivariable}. These controllers often have a proportional-integral-derivative (PID) structure in addition to higher-order filters to compensate parasitic dynamics \citep{steinbuch1998advanced}. Loop-shaping can also be applied to multivariable systems through decoupling or sequential loop closing \citep{oomen2017model}. However, these methods have in common that the design procedure can be difficult as they are based on design rules, insight and experience.

As an alternative, control design based on nonparametric models has been further developed towards automated procedures that utilize FRF measurements to synthesize linear time-invariant (LTI) controllers. At first, these methods were developed along the lines of the classical control theory to synthesize PID controllers \citep{grassi2001integrated}. More recently, these methods have been tailored towards more general control structures that focus on $\mc{H}_\infty$-performance, with many successful applications within the LTI domain \citep{karimi2010fixed,khadraoui2014model}. This was further extended to a framework in which model uncertainties can be incorporated into the control design, such that a robustly stabilizing controller is synthesized to accomodate for the variations in the plant \citep{karimi2007robust,karimi2018robust}. However, this typically comes at the cost of performance.

The paradigm of linear parameter-varying (LPV) systems has been developed to provide a systematic framework for the analysis and design of gain-scheduled controllers for nonlinear systems \citep{shamma1990analysis}. An LPV system is characterized by a linear input-output (IO) map, similar to the LTI framework, where now the dynamics depend on an exogenous time-varying signal whose values can be measured on-line. This so-called scheduling variable $p$ can be used to capture the nonlinear or operating condition-dependent dynamics of a system. Typically, a priori information on the scheduling variable is known, such as the range of variation. The class of LPV systems is supported by a well-developed model-based control and identification theory, with approaches that can be viewed as extensions of LTI control methodologies, see, e.g., \citep{hoffmann2015survey,mohammadpour2012control} and the references therein. Also, data-driven control design techniques in the time-domain exist \citep{formentin2016direct}. With respect to data-driven controller synthesis based on frequency response functions, only a handful of methodologies exist \citep{kunze2007gain,karimi2013hinf,bloemers2019towards_lpv_synthesis}. These methods have in common that an LPV controller is synthesized such that, locally for every operating point, stability and performance can be guaranteed. %The drawback is that these methods are limited to stable systems only and the stability and performance conditions together with the controller parameterization are conservative.

Although data-driven controller synthesis based on FRF data enables systematic design approaches in the LTI framework, within the LPV framework, these are conservative and limited to stable systems only for. Within the LTI literature, necessary and sufficient frequency-domain analysis conditions exist for robust stability \citep{rantzer1994}. These conditions have been used in \citep{karimi2018robust} to synthesize controllers for even unstable LTI systems, guaranteeing stability and $\mc{H}_\infty$-performance. The aim of this paper is to overcome the limitations currently present for data-driven LPV controller synthesis in the frequency-domain by (i) developing necessary and sufficient analysis and synthesis conditions for (possibly) unstable systems and controllers, and (ii), allowing a rational LPV controller parameterization.

The main contributions of this paper are (C1) a procedure to synthesize LPV controllers for possibly unstable single-input single-output plants that achieve local internal stability and $\mc{H}_\infty$-performance guarantees. This is achieved by the following sub-contributions. 
\begin{enumerate}[{C1},left=\parindent]
	\setcounter{enumi}{1}
	\item {\label{Contribution:2}} Development of a local LPV frequency-domain stability analysis condition.
	\item {\label{Contribution:3}} Development of an LPV frequency-domain performance analysis condition.
\end{enumerate}
The results in \citet{rantzer1994} are recovered as a special case for stable systems, constituting to \ref{Contribution:2}. Contribution \ref{Contribution:3} is achieved by developing new insights into the performance conditions presented in \citep{karimi2018robust}, that relate to the robust control theory and consequently to LPV systems by means of the main loop theorem, see e.g., \citep{zhou1996robust}. Furthermore, the results in \citep{karimi2018robust} are recovered as a special case when the scheduling disappears. Finally, C1 is achieved by utilizing a global parameterization of the LPV controller, for which local stability and performance guarantees are provided by means of \ref{Contribution:2} and \ref{Contribution:3}.

The paper is organized as follows. In Section \ref{sec:problemFormulation} the problem setting is defined and the problem of interest is formulated. Then, in Section \ref{sec:analysis_conditions} analysis conditions for stability and performance are derived, constituting to \ref{Contribution:2} and \ref{Contribution:3}. This is followed by the derivation of a synthesis procedure and the main contribution C1 in Section \ref{section:synthesis}. In Section \ref{sec:results}, the capabilities of the proposed methodology are demonstrated by means of a simulation example. Finally, conclusions are drawn in Section \ref{sec:conclusion}. 

Throughout this paper, $\bbR$ denotes the set of real numbers and $\mbb{C}$ is the set of complex numbers. The imaginary axis is denoted by $\mbb{C}_{0}$ and the right half-plane is denoted by $\mbb{C}_{+}$. The real part of a complex number $z \in \mbb{C}$ is denoted by $\Re \{ z \}$. The imaginary unit is denoted by $i = \sqrt{-1}$. The set of real rational proper and stable transfer functions is denoted as $\RHinf$, while the continuous frequency set associated with the Fourier transform is given by $\Omega := \{ \bbR \cup \{ \infty \} \}$.

%===============================================================================
\section{Problem formulation}
\label{sec:problemFormulation}

% ------------------------------------------------------------------------------
\subsection{Preliminaries}
Consider the single-input single-output (SISO), continuous-time (CT) LPV system, with LPV state-space representation \citep{toth2010modeling}:
\begin{align}
	\label{eqn:LPVss}
	G_p:
	\begin{cases}
		\dot{x}(t) &= A(p(t))x(t) + B(p(t))u(t),\\
		y(t) &= C(p(t))x(t) + D(p(t))u(t),
	\end{cases}
\end{align}
where $x : \mbb{R} \rightarrow \mbb{X} \subseteq \mbb{R}^{n_\mrm{x}}$ denotes the state variable, $u : \mbb{R} \rightarrow \mbb{U} \subseteq \mbb{R}$ is the input signal, $y: \mbb{R} \rightarrow \mbb{Y} \subseteq \mbb{R}$ is the output signal and $p : \mbb{R} \rightarrow \mbb{P} \subseteq \mbb{R}^{n_\mrm{p}}$ the scheduling variable.

When the scheduling signal $p(t) \equiv \mrm{p}$ is frozen in time, the scheduling-dependent matrices in \eqref{eqn:LPVss} become time-invariant, i.e., with slight abuse of notation
\begin{equation}
\label{eqn:LPVssFrozen}
	G_\mrm{p} = 
	\begin{pmatrix}
		\begin{array}{c|c}
			A(\mrm{p}) & B(\mrm{p}) \\ \hline
			C(\mrm{p}) & D(\mrm{p})
		\end{array}
	\end{pmatrix}
\end{equation}
represents the LPV system with state-space form \eqref{eqn:LPVss} for constant scheduling $\mrm{p}$. For a given $\mrm{p} \in \mbb{P}$, \eqref{eqn:LPVssFrozen} describes the local behavior of \eqref{eqn:LPVss}. Hence, \eqref{eqn:LPVssFrozen} is referred to as the frozen behavior of \eqref{eqn:LPVss}.

Taking the Laplace transform of \eqref{eqn:LPVssFrozen} with zero initial conditions results in
\begin{equation}
\label{eqn:LPVIOFrozen}
	\hat{y}(s) = \left(C(\mrm{p})(sI - A(\mrm{p}))^{-1}B(\mrm{p}) + D(\mrm{p})\right)\hat{u}(s),
\end{equation}
where $G_\mrm{p}(s) = C(\mrm{p})(sI - A(\mrm{p}))^{-1}B(\mrm{p}) + D(\mrm{p})$ and $s$ is the Laplace variable. The frozen behavior \eqref{eqn:LPVssFrozen} also has a corresponding Fourier transform
\begin{equation}
\label{eqn:fFRF}
	Y(i\omega) = G_\mrm{p}(i\omega)U(i\omega),
\end{equation}
where $i$ is the complex unit, $\omega \in \mbb{R}$ is the frequency and $G_\mrm{p}(i\omega)$ represents the frozen Frequency Response Function (fFRF) of \eqref{eqn:LPVss} for every constant $p(t) \equiv \mrm{p} \in \mbb{P}$ \citep{schoukens2019frequency}.

% ------------------------------------------------------------------------------
%\subsection{Internal stability}
%\label{subsection:internal_stability}
\begin{figure}[t]
	\centering
	\includegraphics[scale=1.00]{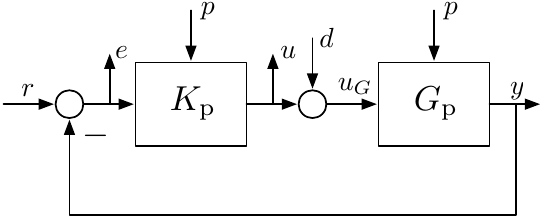}
	\caption{Feedback interconnection, including 4-block shaping problems, depending on the scheduling signal.}
	\label{fig:internal_stability}
\end{figure}

% ------------------------------------------------------------------------------
\subsection{Problem statement}
\label{subsection:problem_statement}
The problem addressed in this paper is to design an LPV controller directly from fFRF measurement data. We denote the data $\mc{D}_{N, \mrm{p}_\tau} = \{ G_\mrm{p}(i\omega_k), \mrm{p}_\tau \}_{k=1}^{N}$, obtained at the set of operating points $\mc{P} = \{ \mrm{p}_\tau \}_{\tau=1}^{N_\mrm{loc}} \subset \mbb{P}$. Consider the feedback interconnection in Figure \ref{fig:internal_stability}. The objective is to design a controller $K_p$ such that the following requirements are satisfied:
\begin{enumerate}[{R1},left=\parindent]
	\item {\label{requirement1}} The closed-loop system in Figure \ref{fig:internal_stability} is internally stable in the local sense for all $p(t) \equiv \mrm{p} \in \mc{P}$.
	\item {\label{requirement2}} The performance channels $(r, d) \mapsto (e, u)$ in Figure \ref{fig:internal_stability} are bounded in the local $\mc{H}_\infty$-norm sense by $\gamma>0$ for all $\mrm{p} \in \mc{P}$.
\end{enumerate}

In the next section, a rational controller parameterization is introduced that allows for a specific formulation of internal stability. This forms the basis to develop analysis conditions for internal stability and $\mc{H}_\infty$-performance. The theory is first formulated for $\mrm{p} \in \mbb{P}$ for the sake of generality. This also ensures \ref{requirement1} and \ref{requirement2} for $\mrm{p} \in \mc{P}$.

%===============================================================================
\section{Stability and performance analysis conditions}
\label{sec:analysis_conditions}
This section presents local LPV stability and performance analysis conditions. This constitutes to requirements \ref{requirement1} and \ref{requirement2} and contributions \ref{Contribution:2} and \ref{Contribution:3}, respectively. Based on these results, a data-driven synthesis procedure is developed. Throughout this section, first the results are presented with a continuous frequency spectrum $\Omega = \{\mbb{R} \cup \{ \infty \} \}$, which will be restricted later by a finite frequency grid $\Omega_N = \{ \omega_k \}_{k=1}^{N}$ corresponding to the data $\mc{D}_{N,\mrm{p}_\tau}$. 

% ------------------------------------------------------------------------------
\subsection{Stability}
Figure \ref{fig:internal_stability} corresponds to the internal stability problem \citep[Chapter 3]{doyle1992feedback}. For a frozen $\mrm{p} \in \mbb{P}$, let the IO map $T(G_\mrm{p}, K_\mrm{p}) : (r, -d) \mapsto (e, u)$ in Figure \ref{fig:internal_stability} be defined by 
\begin{equation}
\label{eqn:four-block}
	T(G_\mrm{p}, K_\mrm{p}) =
	\begin{bmatrix}
		S_\mrm{p} & S_\mrm{p}G_\mrm{p} \\
		K_\mrm{p}S_\mrm{p} & T_\mrm{p}		
	\end{bmatrix}, 	
\end{equation}
with $S_\mrm{p} = (1+G_\mrm{p}K_\mrm{p})^{-1}$ and $T_\mrm{p} = 1-S_\mrm{p}$. If $G_\mrm{p}, K_\mrm{p} \in \RHinf$, then $T(G_\mrm{p}, K_\mrm{p})$ is internally stable if all elements in the IO map $T(G_\mrm{p}, K_\mrm{p})$, defined by \eqref{eqn:four-block}, are stable \citep[Chapter 3]{doyle1992feedback}. If $T(G_\mrm{p}, K_\mrm{p}) \in \RHinf$ holds for all frozen $\mrm{p} \in \mbb{P}$ then the closed-loop LPV system is called locally internally stable. To assess internal stability for unstable $G_\mrm{p}$ or $K_\mrm{p}$, introduce
\begin{equation}
\label{eqn:Gparameterization}
	G_\mrm{p} = N_{G_\mrm{p}}D_{G_\mrm{p}}^{-1}, \quad \{ N_{G_\mrm{p}}, D_{G_\mrm{p}}\} \in \RHinf.
\end{equation}
The two transfer functions $\{ N_{G_\mrm{p}}, D_{G_\mrm{p}}\}$ are a coprime factorization over $\RHinf$ if there exist two other transfer functions $\{ X_\mrm{p}, Y_\mrm{p}\} \in \RHinf$ such that they satisfy the B\'{e}zout identity
\begin{equation}
\label{eqn:bezout}
	N_{G_\mrm{p}} X_\mrm{p} + D_{G_\mrm{p}}Y_\mrm{p} = 1.
\end{equation}
Correspondingly, $K_\mrm{p}$ admits the coprime factorization
\begin{equation}
\label{eqn:Kparameterization}
	K_\mrm{p} = N_{K_\mrm{p}}D_{K_\mrm{p}}^{-1}, \quad \{ N_{K_\mrm{p}}, D_{K_\mrm{p}}\} \in \RHinf.
\end{equation}
Using these definition, \eqref{eqn:four-block} can be represented by
\begin{equation}
\label{eqn:four-block-coprime}
	T(G_\mrm{p}, K_\mrm{p}) = 
	D_{\mrm{p}}^{-1}
	\begin{bmatrix}
		D_{G_\mrm{p}}D_{K_\mrm{p}} & N_{G_\mrm{p}}D_{K_\mrm{p}} \\
		D_{G_\mrm{p}}N_{K_\mrm{p}} & N_{G_\mrm{p}}N_{K_\mrm{p}}
	\end{bmatrix},
\end{equation}
with characteristic equation 
\begin{equation}
\label{eqn:characteristicEquation}
	D_{\mrm{p}} = D_{G_\mrm{p}}D_{K_\mrm{p}} + N_{G_\mrm{p}}N_{K_\mrm{p}}.
\end{equation}
The feedback system in Figure \ref{fig:internal_stability} is internally stable if and only if $D_{\mrm{p}}^{-1} \in \RHinf$. This follows from the B\'ezout identity, i.e., set $N_{K_\mrm{p}} = X_\mrm{p}$ and $D_{K_\mrm{p}} = Y_\mrm{p}$, then the characteristic equation \eqref{eqn:characteristicEquation} equals the B\'{e}zout identity \eqref{eqn:bezout} and the feedback system is internally stable.
 Similarly, the closed-loop LPV system is called locally internally stable if these conditions hold for all $\mrm{p} \in \mbb{P}$.

For the channel transfer $w \mapsto z$, where $w \in \{r, -d\}$ and $z \in \{e,u\}$, let
\begin{equation}
\label{eqn:TSISO_definition}
	T_{z,w}(G_\mrm{p}, K_\mrm{p}) = N_{\mrm{p}}D_{\mrm{p}}^{-1},
\end{equation}
with $\{ N_{\mrm{p}}, D_{\mrm{p}} \} \in \RHinf$ and $T_{z,w}(G_\mrm{p}, K_\mrm{p}) \in \RHinf$, define the corresponding SISO element of \eqref{eqn:four-block-coprime}. For example, $T_{r,e}(G_\mrm{p}, K_\mrm{p}) = N_{\mrm{p}}D_{\mrm{p}}^{-1}$ with $N_{\mrm{p}} = D_{G_\mrm{p}}D_{K_\mrm{p}}$ defines the sensitivity $S_\mrm{p}$ in \eqref{eqn:four-block} and \eqref{eqn:four-block-coprime}.
%\begin{remark}
%	The factorization in \eqref{eqn:four-block-coprime} defines the general case. In the situation that both the plant and controller are stable, \eqref{eqn:four-block} can be recovered by selecting $\{ N_{G_\mrm{p}}, D_{G_\mrm{p}} \} = \{ G_\mrm{p}, 1 \}$ and $\{ N_{K_\mrm{p}}, D_{K_\mrm{p}} \} = \{ K_\mrm{p}, 1 \}$.
%\end{remark}

The following theorem presents analysis conditions to verify internal stability of a closed-loop LPV system locally, given the plant and controller only. 
\begin{theorem}
\label{thm:rantzer1994_coprime}
Let $G_\mrm{p}$ and $K_\mrm{p}$ be as defined in \eqref{eqn:Gparameterization} and \eqref{eqn:Kparameterization}, respectively, and let $D_{\mrm{p}} \in \RHinf$ be as defined in \eqref{eqn:characteristicEquation}. Then the following conditions are equivalent. For all $\mrm{p} \in \mbb{P}$
\begin{enumerate}[{\ref{thm:rantzer1994_coprime}\alph*)},left=\parindent]
	\item {\label{rantzer1994_coprime:stability_coprime}} $D_{\mrm{p}}^{-1} \in \RHinf$.
	\item {\label{rantzer1994_coprime:nonzero}} $D_{\mrm{p}}(s) \neq 0, \, \forall s \in \mbb{C}_+ \cup \mbb{C}_{0} \cup \{ \infty \}$.
	\item {\label{rantzer1994_coprime:posreal_rhinf}} There exists a multiplier $\alpha_\mrm{p}, \alpha_\mrm{p}^{-1} \in \RHinf$ such that
		\begin{equation*}
			\Re \{ D_{\mrm{p}}(i\omega)\alpha_\mrm{p}(i\omega) \} > 0, \, \forall \omega \in \Omega. 
			\vspace{1ex}	% Fix spacing after equation
		\end{equation*} 		
\end{enumerate}
\end{theorem}
The proof can be found in Appendix \ref{appdx:ProofThm1}. Theorem \ref{thm:rantzer1994_coprime} provides an analysis condition to verify local stability for the closed-loop system if instead of a parametric model $N_{G_\mrm{p}}$ and $D_{G_\mrm{p}}$ are only given in terms of local frequency-domain data. The test relates to the Nyquist stability theorem, however without the need to visualize the data in terms of a plot and counting encirclements. Instead, if a transfer function $\alpha_\mrm{p}, \alpha_\mrm{p}^{-1} \in \RHinf$ can be found such that statement \ref{rantzer1994_coprime:posreal_rhinf} holds, then Nyquist stability holds and the system is internally stable. The next subsection presents the extension towards a performance analysis condition.

% ------------------------------------------------------------------------------
\subsection{Performance}
This subsection presents an analysis condition to assess locally the $\mc{H}_\infty$-performance of an LPV system. This constitutes contribution \ref{Contribution:3}. To derive performance analysis conditions, we first present the main loop theorem.

\begin{figure}[]
	\centering
	\vspace{1ex}
	\begin{subfigure}[b]{0.45\columnwidth}
		\centering
		\includegraphics[scale=1.00]{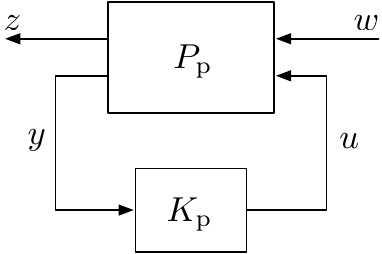}
		\caption{}
		\label{fig:PKForm}
	\end{subfigure}
	\hfill
	\begin{subfigure}[b]{0.45\columnwidth}
		\centering
		\includegraphics[scale=1.00]{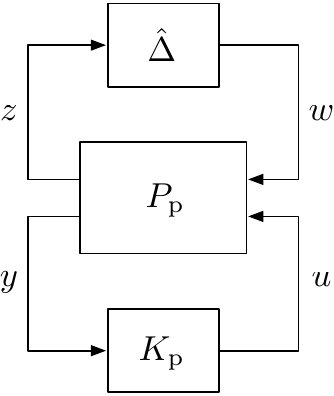}
		\caption{}
		\label{fig:MLBD}
	\end{subfigure}
	\caption{Generalized LPV plant (a); and performance of the SISO closed-loop map $w \mapsto z$ (b).}
	\vspace{-1ex}
\end{figure}

Consider the transfer function $T_{z,w}(G_\mrm{p}, K_\mrm{p}) \in \RHinf$ of interest in Figure \ref{fig:PKForm}, such that $w \mapsto z : T_{z,w}(G_\mrm{p}, K_\mrm{p})$, and let ${ \hat{\Delta} \in \mathbf{B\hat{\Delta}} }$, with
\begin{equation}
\mathbf{B\hat{\Delta}} := \Set{\hat{\Delta} \in \RHinf}{\lvert \hat{\Delta}(i\omega) \rvert < 1, \, \forall \omega \in \Omega}
\end{equation}
a fictitious uncertainty that represents the $\mc{H}_\infty$-performance criterion. Then, $\mc{H}_\infty$-performance of the system in Figure \ref{fig:PKForm} is equivalent to Figure \ref{fig:MLBD} \citep[Theorem 8.7]{skogestad2001MFC}. This is captured by the following theorem.
\begin{theorem}[Main loop theorem]
\label{thm:main_loop}
	Let $W_T \in \RHinf$ and $T_{z,w}(G_\mrm{p}, K_\mrm{p})$ be defined as in \eqref{eqn:TSISO_definition}. The following statements are equivalent. For all $\mrm{p} \in \mbb{P}$
	\begin{enumerate}[{\ref{thm:main_loop}\alph*)},left=\parindent]
		\item {\label{thm:main_loop_a}} $\underset{\omega \in \Omega}{\sup} \, \lvert  W_T(i\omega) T_{z,w}(G_\mrm{p}, K_\mrm{p})(i\omega) \rvert \leq \gamma$.
		\item {\label{thm:main_loop_b}} $ \begin{aligned}[t] 1 - \gamma^{-1}W_T(i\omega) T_{z,w}(G_\mrm{p}, K_\mrm{p})(i\omega)\hat{\Delta}(i\omega) \neq 0, \\ \forall \omega \in \Omega, \, \forall \hat{\Delta} \in \mathbf{B\hat{\Delta}}. \end{aligned}$
	\end{enumerate}
\end{theorem}
Theorem \ref{thm:main_loop} is a special case of \citet[Theorem 11.7]{zhou1996robust}, where the weighting filter $W_T$ is introduced to specify the frequency-dependent design requirements on the map $w \mapsto z$. The theorem connects nominal performance to robust stability through the interconnection of the performance channels with a fictitious uncertainty block, see Figure \ref{fig:MLBD}.

The main loop theorem provides useful insight into performance. In the data-driven setting, the absence of a parametric model of $T_{z,w}(G_\mrm{p}, K_\mrm{p})$ makes it difficult to turn statement \ref{thm:main_loop_b} into a convex constraint as it is generally done in model-based LPV synthesis approaches for gain-scheduling \citep{hoffmann2015survey}. Hence, in that case statement \ref{thm:main_loop_b} is needed to be evaluated for an infinite set of realizations of the fictitious uncertainty $\hat{\Delta}$, for example, as in \citep{van2018frequency}. The contribution in this paper is to utilize Theorem \ref{thm:rantzer1994_coprime} together with Theorem \ref{thm:main_loop} to derive a single theorem to analyze both stability and performance without the need to sample $\hat{\Delta}$.
\begin{theorem}
\label{thm:performance_analysis}
	Let $W_T \in \RHinf$ and $T_{z,w}(G_\mrm{p}, K_\mrm{p})$ be defined as in \eqref{eqn:TSISO_definition}. Requirements \ref{requirement1} and \ref{requirement2} are satisfied if and only if there exists a multiplier $\alpha_\mrm{p} \in \RHinf$ with $\alpha_\mrm{p}^{-1} \in \RHinf$  such that
	\begin{equation}
	\begin{split}
	\label{eqn:performance_analysis}
		\Re \{ (D_{\mrm{p}}(i\omega) - \gamma^{-1}\lvert W_T(i\omega)N_{\mrm{p}}(i\omega) \rvert)\alpha_\mrm{p}(i\omega) \} > 0, \\
		\forall \omega \in \Omega, \, \forall \mrm{p} \in \mbb{P}.
	\end{split}
	\end{equation} 
\end{theorem}

\begin{figure}[]
	\centering
	\includegraphics[scale=1.00]{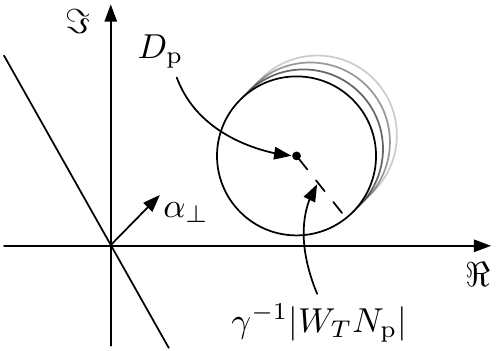}
	\caption{Illustration of stability and $\mc{H}_\infty$-performance. The transfer function $\alpha_\mrm{p}$ represents, for each frequency, a line passing through the origin. If this line does not intersect with the disks $D_{\mrm{p}} - \gamma^{-1}\lvert W_TN_{\mrm{p}} \lvert$ for each frequency, then the disks exclude the origin and (2b) must hold.}
	\label{fig:performance_illustration}
\end{figure}

The proof is given in Appendix \ref{appdx:ProofThm3}. Theorem \ref{thm:performance_analysis} states that the performance condition \ref{thm:main_loop_a} is satisfied if and only if for each frequency $\omega \in \Omega$ and scheduling value $\mrm{p} \in \mbb{P}$ the disks with radius $\gamma^{-1}\lvert W_T N_{\mrm{p}} \rvert$, centered at $D_{\mrm{p}}$, do not include the origin. This holds if there exists a transfer function $\alpha_\mrm{p}, \alpha_\mrm{p}^{-1} \in \RHinf$, representing for each frequency a line passing through the origin, that does not intersect with the disks. This is illustrated in Figure \ref{fig:performance_illustration}. Theorem \ref{thm:performance_analysis} also implies internal stability because $\Re \{ D_\mrm{p}(i\omega)\alpha(i\omega)\} > 0$ implies internal stability by Theorem \ref{thm:rantzer1994_coprime}. 

The analysis condition is especially useful as it provides a local stability and performance result given only a controller and the data $\mc{D}_{N,\mrm{p}_\tau}$. Similar to the stability analysis condition, a parametric model is not required.
%\begin{figure}[]
%	\centering
%	\includegraphics[scale=1.00]{Figures/performance_drawing2}
%	\caption{Illustration of stability and $\mc{H}_\infty$-performance. The transfer function $\alpha_\mrm{p}$ represents, for each frequency, a line passing through the origin. If this line does not intersect with the disks $D_{\mrm{p}} - \gamma^{-1}\lvert W_TN_{\mrm{p}} \lvert$ for each frequency, then the disks exclude the origin and (\ref{thm:main_loop_b} must hold.}
%	\label{fig:performance_illustration}
%\end{figure}

% ------------------------------------------------------------------------------
\subsection{Synthesis}
We give an equivalent formulation of Theorem \ref{thm:performance_analysis} that is useful for controller synthesis.
\begin{theorem}
\label{thm:synthesis}
	Given $G_\mrm{p} = N_{G_\mrm{p}}D_{G_\mrm{p}}^{-1}$, with $\{ N_{G_\mrm{p}}, D_{G_\mrm{p}} \} \in \RHinf$ coprime, as defined in \eqref{eqn:Gparameterization}, and a weighting filter $W_T \in \RHinf$, the following statements are equivalent.
	\begin{enumerate}[{\ref{thm:synthesis}\alph*)},left=\parindent]
		\item {\label{thm:karimi_a}} There exists a proper rational controller $K_\mrm{p}$ that achieves internal stability and performance as defined in requirements \ref{requirement1} and \ref{requirement2}, respectively.
		\item {\label{thm:karimi_b}} There exists a controller $K_\mrm{p} = N_{K_\mrm{p}}D_{K_\mrm{p}}^{-1}$, with $\{ N_{K_\mrm{p}}, D_{K_\mrm{p}} \} \in \RHinf$, as defined in \eqref{eqn:Kparameterization}, such that
			\begin{equation}
			\begin{split}
			\label{eqn:synthesis_constraint}
				\Re \{ D_{\mrm{p}}(i\omega) \} > \gamma^{-1}\lvert W_T(i\omega)N_{\mrm{p}}(i\omega) \rvert, \\
				\forall \omega \in \Omega,  \, \forall \mrm{p} \in \mbb{P}.
			\end{split}
			\end{equation}
	\end{enumerate}
\end{theorem}

The proof can be found in Appendix \ref{appdx:ProofThm4}. Theorem \ref{thm:synthesis} is the main result in this paper and presents a local $\mc{H}_\infty$-optimal controller synthesis condition given only data $\mc{D}_{N,\mrm{p}_\tau}$. This is further developed in Section \ref{section:synthesis}, where an optimization problem is formulated and the controller parameterization is discussed.

%\TB{
%\begin{remark}
%	Note that an important outcome of Theorem \ref{thm:synthesis} is that the multiplier $\alpha_\mrm{p}$ can be absorbed into the controller as $\gamma^{-1} \lvert W_T(i\omega)N_{\mrm{p}}(i\omega) \alpha_\mrm{p}(i\omega) \rvert \implies \Re \{ \gamma^{-1} \lvert W_T(i\omega)N_{\mrm{p}}(i\omega) \rvert \alpha_\mrm{p}(i\omega) \}$. Note that the absorbed multiplier changes the considered $N_{\mrm{p}}$ and $D_{\mrm{p}}$, but $\alpha_\mrm{p}$ cancels out in both terms when $K_\mrm{p} = N_{K_\mrm{p}}D_{K_\mrm{p}}^{-1}$ is computed. The price to be paid for this absorption is the increased order of $N_{K_\mrm{p}}$ and $D_{K_\mrm{p}}$.
%\end{remark}
%\begin{remark}
%	\citep[Theorem 1]{karimi2018robust} is recovered in the special case when the plant and controller are LTI . 
%\end{remark}
%}

%===============================================================================
\section{Controller synthesis}
\label{section:synthesis}
In this section we develop a procedure to synthesize LPV controllers directly from the frequency-domain measurement data $\mc{D}_{N, \mrm{p}_\tau}$. First, an optimization problem is set up in Section \ref{subsection:synthesisOptimization} that characterizes the synthesis problem based on Theorem \ref{thm:synthesis}. This is followed by a discussion on the controller parameterization in Section \ref{subsection:Kparameterization}.

% ------------------------------------------------------------------------------
\subsection{Controller synthesis}
\label{subsection:synthesisOptimization}
Given the data $\{\mc{D}_{N, \mrm{p}_\tau}, \mrm{p}_\tau \in \mc{P}\}$ and a controller parameterization $K_\mrm{p} = N_{K_\mrm{p}}D_{K_\mrm{p}}^{-1}$, the following optimization problem is formulated to satisfy Requirements \ref{requirement1} and \ref{requirement2}:
\begin{equation}
\label{eqn:synthesis}
\begin{aligned}
	& \underset{\theta, \gamma}{\text{min}}
	& & \gamma \\
	& \text{s.t.}
	& & \gamma \Re \{ D_{\mrm{p}}(i\omega, \theta) \} > \lvert W_T(i\omega)N_{\mrm{p}}(i\omega, \theta) \rvert \\
	& & & \forall \omega \in \Omega, \, \mrm{p} \in \mc{P},
\end{aligned}
\end{equation}
where $\theta$ are the controller parameters. 

The optimization problem \eqref{eqn:synthesis} is generally non-convex. However, a linear parameterization of $\{N_{K_\mrm{p}},D_{K_\mrm{p}}\}$ results in a quasi-convex form of \eqref{eqn:synthesis} in the controller parameters $\theta$ and the performance indicator $\gamma$. A bisection algorithm can be used to solve the quasi-convex program. This results in an iterative approach, where for every fixed value of $\gamma$ a second-order cone program is solved. 

To provide stability and performance guarantees, the constraints in \eqref{eqn:synthesis} need to be satisfied on the infinite set $\omega \in \Omega$, leading to a semi-infinite program. One solution is to solve \eqref{eqn:synthesis} for a finite grid of frequencies $\Omega_N = \{ \omega_k \}_{k=1}^{N} \subset \Omega$. The frequencies in this grid have to be chosen dense enough such that a Nyquist curve can be interpreted from the data.

% ------------------------------------------------------------------------------
\subsection{Controller parameterization}
\label{subsection:Kparameterization}
In Section \ref{sec:analysis_conditions} the rational controller factorization is introduced. This section presents the controller parameterization and the requirements that are need to be satisfied.
\begin{enumerate}[label=\roman*)]
	\item The controller must admit the factorization \eqref{eqn:Kparameterization}. This enables tuning of both the poles and zeros of the controller, in contrast to previous data-driven frequency-domain LPV tuning methods \citep{kunze2007gain,karimi2013hinf,bloemers2019towards_lpv_synthesis}.
	\item The scheduling-dependency must be chosen such that $\{ N_{K_\mrm{p}}, D_{K_\mrm{p}}\} \in \RHinf$ for all $\mrm{p} \in \mbb{P}$. 
	\item A linear parameterization of $N_{K_p}$ and $D_{K_p}$ is preferred to keep \eqref{eqn:synthesis} quasi-convex.
	\item The controller structure must be such that the multiplier $\alpha_\mrm{p}$ can be absorbed. This requirement can be alleviated, but consequently results in a bi-linear optimization problem between the controller parameters and multiplier $\alpha_\mrm{p}$. 
	\item A monic structure of $D_{K_\mrm{p}}$ avoids a trivial solution to \eqref{eqn:synthesis}. Furthermore, this ensures that $D_{K_\mrm{p}}^{-1}$ is well-defined for all $\mrm{p} \in \mbb{P}$.
\end{enumerate}

An orthonormal basis function (OBF)-based representation \citep{toth2010modeling} is a natural choice to parameterize the controller factors
\begin{subequations}
\begin{align}
\label{eqn:Nparameterization}
	N_{K_\mrm{p}}(s) &= \sum_{i=0}^{n_N} w_i(\mrm{p})\phi_i(s), \\
	\label{eqn:Dparameterization}
	D_{K_\mrm{p}}(s) &= \sum_{i=0}^{n_D} v_i(\mrm{p})\varphi_i(s),
\end{align}
\end{subequations}
such that the requirements (i)-(v) are satisfied. Here, $\{ \phi_i \}_{i=0}^{n_N}$ and $\{ \varphi_i \}_{i=0}^{n_D}$ with $\phi_0 = \varphi_0 = 1$ and $n_D \geq n_N$ are the sequence of basis functions, with coefficient functions 
\begin{equation}
	w_i(\mrm{p}) = \sum_{\ell=1}^m \breve{w}_{i}^{\ell} \psi_{\ell}(\mrm{p}),
\end{equation}
and similarly for $v_i(p)$. Here, the coefficient functions are formed through a chosen functional dependence, e.g., affine, polynomial or rational dependence characterized by the basis functions $\{ \psi_\ell \}_{\ell=1}^{m}$. See \citep[Chapter 9.2]{toth2010modeling} for an overview of OBF based LPV model structures and their properties.
\begin{remark}
	The concept in this paper is to shape the global behavior of the controller by tuning the parameter-dependent coefficient functions based on their local behavior, i.e., for constant $p$.
\end{remark}

%\begin{figure}[]
%	\centering
%	\includegraphics[scale=1.00]{OBFWiener.pdf}
%	\caption{Input-output graph of the Wiener LPV OBF structure.}
%	\label{fig:WienerOBF}
%	\vspace{-1em}
%\end{figure}
%
%The parameterization of $N_{K_p}$ and $D_{K_p}$ can be viewed as a bank of OBFs, whose output is combined though parameter-dependent coefficient functions, see Figure \ref{fig:WienerOBF}. In this paper, Wiener LPV OBF structures are considered for each factor, which can be represented in the state-space forms:
%\begin{subequations}
%\begin{align}
%\label{eqn:Nss}
%	N_{K_p} &= 
%	\begin{pmatrix}
%		\begin{array}{c|c}
%			A_N & B_N \\ \hline
%			C_N(p(t)) & D_N(p(t))
%		\end{array}
%	\end{pmatrix},
%	\\
%\label{eqn:Dss}
%	D_{K_p} &= 
%	\begin{pmatrix}
%		\begin{array}{c|c}
%			A_D & B_D \\ \hline
%			C_D(p(t)) & D_D(p(t))
%		\end{array}
%	\end{pmatrix}.
%\end{align}
%\end{subequations}
%Equations \eqref{eqn:Nparameterization}-\eqref{eqn:Dss} reveal that requirements (i)-(iv) are satisfied. Requirement (v) is satisfied, w.l.o.g. by $\{ \breve{v}_{i}^{\ell} \}_{i=0}^{\ell=1,\dots,m} = \{1,0, \dots, 0 \}$, or equivalently by setting $D_D(p(t)) \equiv 1$ in \eqref{eqn:Dss}. Because the set of bases is complete w.r.t. $\mc{H}_2$, hence any solution including the optimal solution of \eqref{eqn:synthesis} can be found via parameterizations \eqref{eqn:Nparameterization}--\eqref{eqn:Dparameterization} \cite{karimi2018robust}.
\subsection{Controller implementation}
\label{subsection:Kimplementation}
\begin{figure}[t]
	\centering
	\vspace{1ex}
	\begin{subfigure}[]{0.60\columnwidth}
		\centering
		\includegraphics[scale=1.00]{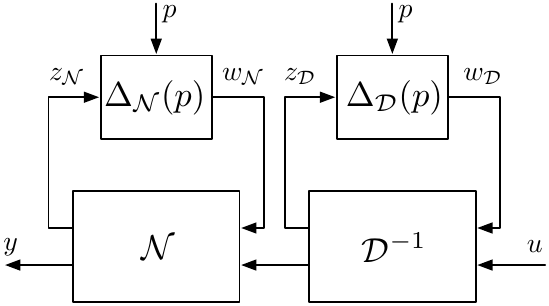}
		\caption{}
		\label{fig:KLFR1}
	\end{subfigure}
	\hfill
	\begin{subfigure}[]{0.37\columnwidth}
		\centering
		\includegraphics[scale=1.00]{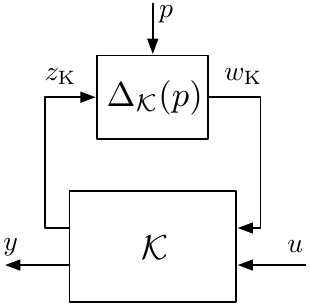}
		\caption{}
		\label{fig:KLFR2}
	\end{subfigure}
	\vspace{-1ex}
	\caption{Controller realization through (a) the series connection of the LFR representations $N_{K_p} = \mc{F}_u(\mc{N}, \Delta_\mc{N}(p))$ and $D_{K_p} = \mc{F}_u(\mc{D}^{-1}, \Delta_\mc{D}(p))$ and (b) $K_p = \mc{F}_u(\mc{K}, \Delta_\mc{K})$, with $\Delta_\mc{K} = \mrm{diag}(\Delta_\mc{N},\Delta_\mc{D})$.}
\label{fig:KLFR}	
\end{figure}

The OBF parameterizations admit a linear fractional representation (LFR)-structure. In this structure, the dependency on the scheduling variable $p$ is extracted by formulating \eqref{eqn:Nparameterization} and \eqref{eqn:Dparameterization} in terms of LTI systems, denoted by $\mc{N}$ and $\mc{D}$, such that $N_{K_p} = \mc{F}_u(\mc{N}, \Delta_\mc{N}(p))$ and $D_{K_p} = \mc{F}_u(\mc{D}^{-1}, \Delta_\mc{D}(p))$, respectively, where $\mc{F}_u$ is the upper linear fractional transformation \cite{zhou1996robust}, see Figure \ref{fig:KLFR1}. As a consequence of controller restriction (v), the inverse input-output map $\mc{D}^{-1}$ exists for all $p \in \mbb{P}$. This inverse is obtained through partial inversion of the IO map, see, e.g., \cite[Chapter 10]{zhou1996robust}. The controller is formed through the series connection of the LFRs $\mc{N}$ and $\mc{D}^{-1}$, resulting in the LFR $\mc{K}$ such that $K_p = \mc{F}_u(\mc{K}, \mrm{diag}(\Delta_\mc{N}, \Delta_\mc{D}) )$, see Figure \ref{fig:KLFR2}.

%===============================================================================
\section{Results}
\label{sec:results}
Consider a DC motor with mass imbalance corresponding to the following dynamic behavior
\begin{subequations}
\label{eqn:unbalanced_disk}
\begin{align}
	\begin{bmatrix}
		\dot{\theta}(t) \\ \ddot{\theta}(t) \\ \dot{I}(t)
	\end{bmatrix}
	&= 
	\begin{bmatrix}
		0 & 1  & 0 \\
		\frac{Mgl}{J}p(t) & -\frac{b}{J} & \frac{K}{J} \\
		0 & -\frac{K}{L} & -\frac{R}{L} 
	\end{bmatrix}
	\begin{bmatrix}
		\theta(t) \\ \dot{\theta}(t) \\ I(t)
	\end{bmatrix}
	 + \begin{bmatrix}
	 	0 \\ 0 \\ \frac{1}{L}
	 \end{bmatrix}
	 u(t), \\
	 y(t) &= \theta(t),
 \end{align}
\end{subequations}
where $\theta \in [-\pi, \pi] = \mbb{Y}$ denotes the rotation angle of the disk and $u \in \mbb{U}$ is the input voltage. Furthermore, we define by $p(t) = \mathrm{sinc}(\theta(t)) \in \mbb{P}$ the scheduling variable. The parameters of the unbalanced disk are given in Table \ref{table:DCnomenclature}. The unbalanced disk, intrinsically an unstable system, can be thought of as an inverted pendulum rotating around its origin. A set of FRF data of the coprime factors $\{N_{G_\mrm{p}}, D_{G_\mrm{p}} \}$ (derived analytically) is obtained at $N_\mrm{loc} = 9$ equidistantly distributed frozen operating points $p \in \mc{P} \subset \mbb{P}$ and $N_\omega = 400$ logarithmically spaced frequency points $\omega \in \Omega_N \subset [10^{-2}, 200\pi]$ rad/s. The data is obtained in a discrete-time setting under a zero-order-hold assumption at a sampling-rate $T_s = 0.005$ sec.
\begin{table}[]
\centering
\caption{Parameters of the unbalanced disk.}
\label{table:DCnomenclature}
\begin{tabular}{@{}l|c|l|l@{}}
\toprule
Parameter                   & 		& Value 				& Unit               	\\ \midrule
Motor torque constant       & $K$ 	& $0.0536$ 				& $\mathrm{Nm/A}$ 		\\
Motor resistance            & $R$ 	& $9.50$ 				& $\Omega$             	\\
Motor impedance             & $L$ 	& $0.84\cdot10^{-3}$	& H      				\\
Disk inertia                & $J$ 	& $2.2\cdot10^{-4}$ 	& $\mathrm{Nm}^2$   	\\
Viscous friction            & $b$ 	& $6.6\cdot10^{-5}$ 	& $\mathrm{Nms/rad}$ 	\\
Additional mass             & $M$ 	& $0.07$ 				& kg                 	\\
Mass - center disk distance & $l$ 	& $0.042$ 				& m                 	\\ \bottomrule
\end{tabular}
\end{table}

\begin{figure}[t]
	\centering
	\vspace{1ex}
	\includegraphics[scale=1.00]{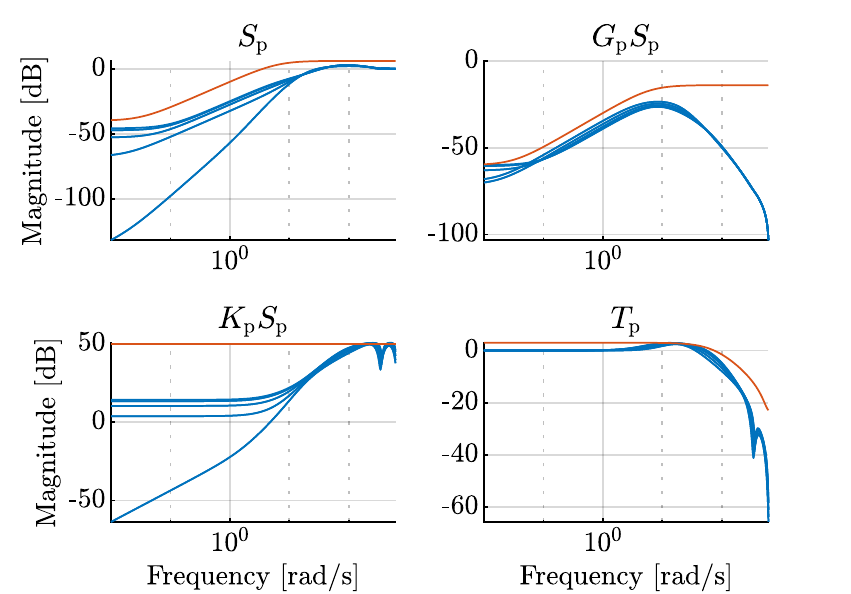}
		\vspace{-3ex}
	\caption{Magnitude plots of the frozen closed-loop sensitivity functions in blue. Their respective weighting filters are shown in orange.}
	\label{fig:bodeweights}
\end{figure}

The control objective is to design a discrete-time controller that achieves good reference tracking and disturbance rejection. The chosen control architecture is that of Figure \ref{fig:internal_stability}, i.e., a four-block problem. The performance specifications are captured in terms of the weighting filters $W_T$, which are shown in Figure \ref{fig:bodeweights}.
The controller factors $\{ N_{K_\mrm{p}}, D_{K_\mrm{p}}\}$ are parameterized by 5th order pulse basis functions, i.e., $\{\phi_i(s)\}^{n_N=5}_{i=0} = \{\varphi_i(s)\}^{n_D=5}_{i=0} =\{z^{-i}\}$. The controller coefficients are chosen to have affine dependence on $p$, resulting in $\{\psi(p) \}_{\ell=1}^{m = 2} = \{ 1, p\}$. 

\begin{figure}[t]
	\centering
	\includegraphics[scale=1.00]{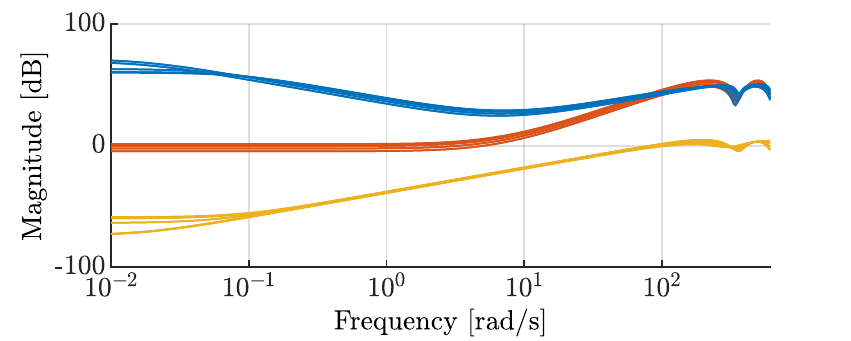}
	\vspace{-3ex}
	\caption{Magnitude plot of the controller $K$ in blue and $\{ N_{K_\mrm{p}}, D_{K_\mrm{p}} \}$, each being parameterized with 5th order pulse bases, in orange and yellow, respectively. The plots are shown at frozen operating points $\mrm{p} \in \mc{P}$.}
	\label{fig:magnitude_controller}
	\vspace{-1ex}
\end{figure}

\begin{figure}[t]
	\centering
	\vspace{1ex}
	\includegraphics[scale=1.00]{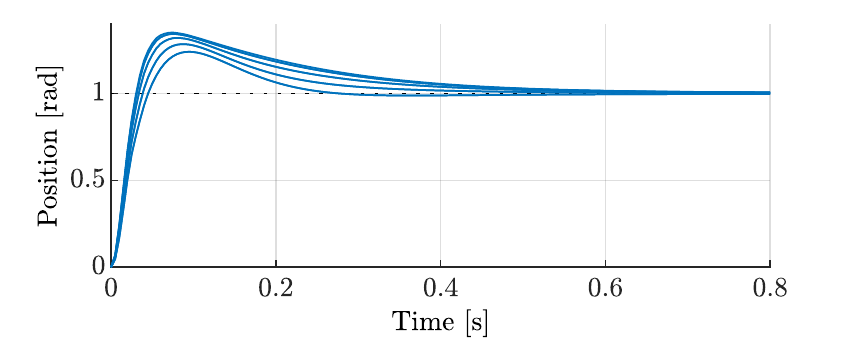}
	\vspace{-3ex}
	\caption{Step responses of the closed-loop system using the data-driven LPV control design. The step responses are shown at frozen operating points $\mrm{p} \in \mc{P}$.}
	\label{fig:step}
	\vspace{-1ex}
\end{figure}

The controller design results in an LPV controller that achieves a performance of $\gamma = 1.15$. The controller parameters are given in Table \ref{table:K_params} and the magnitude plots of the controller and its factorization are given in Figure \ref{fig:magnitude_controller}. The local step responses in Figure \ref{fig:step} shows satisfactory performance, indicating that the LPV controller is able to adapt itself to the operating condition changes of the system based on the available information of the scheduling signal. 

\begin{figure}[t]
	\vspace{-2ex}
	\centering
	\includegraphics[scale=1.00]{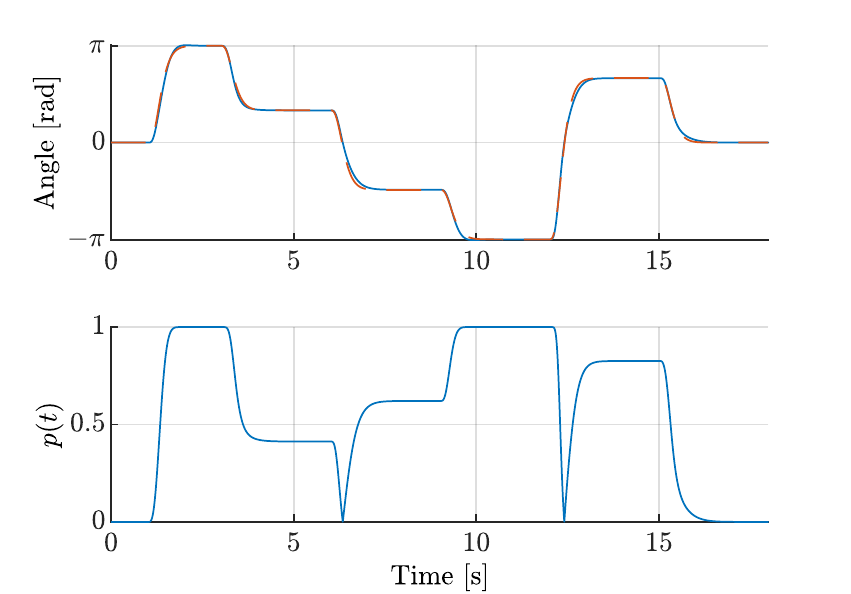}
	\vspace{-3ex}
	\caption{Simulation of the closed-loop nonlinear system with the designed LPV controller. The top plot shows the reference trajectory and angle of the disk in dashed orange and blue, respectively. The bottom plot shows the variation of the scheduling variable.}
	\label{fig:nlsim}
\end{figure}

Figure \ref{fig:nlsim} shows the reference tracking performance of the closed-loop nonlinear system with the designed LPV controller. Remark that, in contrast to before, the scheduling variable is varying over time. It can be observed that stability as well as good performance in terms of reference tracking is achieved for time-varying scheduling trajectories. However, due to the considered local stability and performance setting in this paper, stability can only be guaranteed for sufficiently slow variations of the scheduling parameter.

\begin{table}[]
\centering
\caption{Controller parameters of $\{N_{K_\mrm{p}}, D_{K_\mrm{p}} \}$}
\vspace{-1ex}
\label{table:K_params}
\begin{tabular}{@{}l|rrrrrr@{}}
\toprule
${}_\ell\backslash {}^i$ & 0 & 1 & 2 & 3 & 4 & 5\\ \midrule
$\breve{w}_{i}^{\ell=1}$ & 143.74 & -113.36 & -24.37 & -40.16 & -72.00 & 106.74 \\
$\breve{w}_{i}^{\ell=2}$ & 74.97 & -6.25 & -72.88 & -44.02 & -6.82 & 55.59 \\ \midrule
$\breve{v}_{i}^{\ell=1}$ &  1 & -0.51 & -0.017 & -0.24 & -0.19 & -0.049 \\
$\breve{v}_{i}^{\ell=2}$ &  0 & 0.39 & -0.25 & -0.13 & -0.25 & 0.24 \\ \bottomrule
%1 &  -171.7521 & 342.3863 &  -171.7550 \\ \bottomrule
\end{tabular}
\vspace{-1ex}
\end{table}

%===============================================================================
\section{Conclusion}
\label{sec:conclusion}
This paper presents an LPV controller synthesis approach which enables the design of operating condition-dependent controllers directly from frequency-domain measurement data. This approach enables the design of rational LPV controllers, in contrast to existing data-driven methods in the literature. The capabilities of this approach are presented through a case study on an unstable nonlinear system. We emphasize that only estimates of frozen frequency response functions of the plant are required, no parametric plant model is needed.
%===============================================================================
%\begin{ack}
%Place acknowledgments here.
%\end{ack}

%===============================================================================
\bibliography{LPVS2021}             % bib file to produce the bibliography

%===============================================================================
\appendix
\section{Proof of Theorem \ref{thm:rantzer1994_coprime}}
\label{appdx:ProofThm1}
For a proof of equivalence between \ref{rantzer1994_coprime:stability_coprime} and \ref{rantzer1994_coprime:nonzero}, see \citep[Chapter 3]{doyle1992feedback}. Regarding the equivalence between \ref{rantzer1994_coprime:stability_coprime} and \ref{rantzer1994_coprime:posreal_rhinf} for all $\mrm{p} \in \mbb{P}$, note the following reasoning:

$(\Rightarrow)$ Assume \ref{rantzer1994_coprime:stability_coprime} and let $Q = D_{\mrm{p}}^{-1}$. This implies that the B\'ezout identity \eqref{eqn:bezout} is satisfied for $X_\mrm{p} = N_{K_\mrm{p}}Q$ and $Y_\mrm{p} = D_{K_\mrm{p}}Q$ because, $\{X_\mrm{p}, Y_\mrm{p}\}$ are coprime iff $Q, Q^{-1} \in \RHinf$. Hence, \ref{rantzer1994_coprime:posreal_rhinf} is satisfied by setting $\alpha_\mrm{p} = Q$ because $\Re \{ N_{G_\mrm{p}} X_\mrm{p} + D_{G_\mrm{p}}Y_\mrm{p} \} = 1$ for all  $\omega \in \Omega$.

$(\Leftarrow)$ Assume \ref{rantzer1994_coprime:posreal_rhinf} and let $V = D_{\mrm{p}}\alpha_\mrm{p}$. Note that $V, V^{-1} \in \RHinf$ because \ref{rantzer1994_coprime:posreal_rhinf} implies that $D_{\mrm{p}}\alpha_\mrm{p}$ is bi-proper and has no right half-plane (RHP) zeros. Then $D_{\mrm{p}} = V\alpha_\mrm{p}^{-1}$ satisfies the B\'{e}zout identity \eqref{eqn:bezout}, therefore $D_{\mrm{p}}^{-1} \in \RHinf$. Thus \ref{rantzer1994_coprime:posreal_rhinf} implies \ref{rantzer1994_coprime:stability_coprime} and consequently \ref{rantzer1994_coprime:nonzero}. This completes the proof.

\section{Proof of Theorem \ref{thm:performance_analysis}}
\label{appdx:ProofThm3}
Requirement \ref{requirement2} can be equivalently stated using Theorem \ref{thm:main_loop}, Condition \ref{thm:main_loop_b}, i.e.,
	\begin{equation}
	\begin{split}
	\label{eqn:proof_performance_analysis_1}
			1 - \gamma^{-1}W_T(i\omega)T_{z,w}(G_\mrm{p}, K_\mrm{p})(i\omega)\hat{\Delta}(i\omega) \neq 0, \\ 
			\forall \omega \in \Omega, \, \forall \mrm{p} \in \mbb{P}, \, \forall \hat{\Delta} \in \mathbf{B\hat{\Delta}}.
	\end{split}
	\end{equation}
	As $D_\mrm{p} \in \RHinf$, $D_\mrm{p}(i\omega)\neq 0,$ $\forall \omega \in \Omega$ and by multiplying \eqref{eqn:proof_performance_analysis_1} with it, the resulting non-singularity condition is:
	\begin{equation}
	\begin{split}
	\label{eqn:proof_performance_analysis_1b}
			D_{\mrm{p}}(i\omega) - \gamma^{-1} W_T(i\omega)N_{\mrm{p}}(i\omega) \hat{\Delta}(i\omega)\neq 0, \\ 
			\forall \omega \in \Omega, \, \forall \mrm{p} \in \mbb{P}, \, \forall \hat{\Delta} \in \mathbf{B\hat{\Delta}}.
	\end{split}
	\end{equation}
	Based on a homotopy argument, \eqref{eqn:proof_performance_analysis_1b} corresponds to Condition 1b) in Theorem \ref{thm:rantzer1994_coprime}, which through 1c) is equivalent with
%	\begin{equation}
%	\label{eqn:proof_performance_analysis_2}
%	\begin{split}
%		\Re \{ (1 - \gamma^{-1}W_T(i\omega)T_{z,w}(G_\mrm{p}, K_\mrm{p})(i\omega)\hat{\Delta}(i\omega))\alpha_\mrm{p}(i\omega) \} > 0, \\ 
%		\forall \omega \in \Omega, \, \forall \mrm{p} \in \mbb{P}, \, \hat{\Delta} \in \mathbf{B\hat{\Delta}}.
%	\end{split}
%	\end{equation}
%	Substituting $\eqref{eqn:TSISO_definition}$ for $T_{z,w}(G_\mrm{p}, K_\mrm{p})$ in \eqref{eqn:proof_performance_analysis_2} and multiplying with $D_\mrm{p}$ results in
	\begin{equation}
	\label{eqn:proof_performance_analysis_3}
	\begin{split}
		\Re \{ (D_{\mrm{p}}(i\omega) - \gamma^{-1} W_T(i\omega)N_{\mrm{p}}(i\omega) \hat{\Delta}(i\omega))\alpha_\mrm{p}(i\omega) \} \! > \! 0, \\ 
		\forall \omega \in \Omega, \, \forall \mrm{p} \in \mbb{P}, \, \hat{\Delta} \in \mathbf{B\hat{\Delta}}.
	\end{split}
	\end{equation}
%	Note
%	that multiplying with $D_\mrm{p}$ is not going to change the condition.
	When $\hat{\Delta} = 0 \in \mathbf{B\hat{\Delta}}$, \eqref{eqn:proof_performance_analysis_3} reduces to $\Re \{ D_{\mrm{p}}(i\omega)\alpha_\mrm{p}(i\omega) \} > 0$, which is the same as Condition \ref{rantzer1994_coprime:posreal_rhinf} in Theorem \ref{thm:rantzer1994_coprime}, hence \eqref{eqn:proof_performance_analysis_3} implies requirement \ref{requirement1}.
	
	Let $1\geq \epsilon>0$ and consider  \eqref{eqn:proof_performance_analysis_3} on
	\begin{equation}
\mathbf{B}_\epsilon \mathbf{\hat{\Delta}} := \Set{\hat{\Delta} \in \RHinf}{\lvert \hat{\Delta}(i\omega) \rvert \leq 1-\epsilon, \, \forall \omega \in \Omega},
\end{equation}
	which is the scaled closed uncertainty ball contained in $\mathbf{B\hat{\Delta}}$. Since any $\hat{\Delta} \in \mathbf{B}_\epsilon \mathbf{\hat{\Delta}}$ represents a rotation and contraction in the complex plane, it is necessary and sufficient to check \eqref{eqn:proof_performance_analysis_3} on the boundary only, i.e., for $\hat{\Delta} \in \partial  \mathbf{B}_\epsilon \mathbf{\hat{\Delta}}$, with $\lvert \hat \Delta(i\omega) \rvert = 1-\epsilon$, $\forall \omega \in \Omega$. 
Note that, in \eqref{eqn:proof_performance_analysis_3}, $W_T(i\omega)N_{\mrm{p}}(i\omega)$ only represents complex scaling of this ball which is centered at $D_{\mrm{p}}(i\omega)$. Hence, \eqref{eqn:proof_performance_analysis_3} restricted on $\mathbf{B}_\epsilon \mathbf{\hat{\Delta}}$ is equivalent with 
	\begin{equation}
	\begin{split}
		\Re \{ (D_{\mrm{p}}(i\omega) - \gamma^{-1}(1-\epsilon)\lvert W_T(i\omega)N_{\mrm{p}}(i\omega) \rvert)\alpha_\mrm{p}(i\omega) \} > 0, \\
		\forall \omega \in \Omega, \, \forall \mrm{p} \in \mbb{P}.
	\end{split} \label{scaled_eq}
	\end{equation} 
	This means that if \eqref{scaled_eq} holds, then violation of \eqref{eqn:proof_performance_analysis_3} can only happen in $\mathbf{B\hat{\Delta}} \setminus \mathbf{B}_\epsilon \mathbf{\hat{\Delta}}$. As \eqref{scaled_eq} is continuous in $\epsilon$, by taking the limit $\epsilon \rightarrow 0$, 
	$\mathbf{B\hat{\Delta}} \setminus \mathbf{B}_\epsilon \mathbf{\hat{\Delta}}\rightarrow \emptyset$ and we obtain that \eqref{eqn:performance_analysis} is equivalent with \eqref{eqn:proof_performance_analysis_3}. This completes the proof.

\section{Proof of Theorem \ref{thm:synthesis}}
\label{appdx:ProofThm4}
$(\Rightarrow)$ Assume $K_\mrm{p} = \tilde{N}_{K_\mrm{p}} \tilde{D}_{K_\mrm{p}}^{-1}$ satisfies \ref{thm:karimi_a}. Then, by Theorem \ref{thm:performance_analysis}, there exists an $\alpha_\mrm{p}, \alpha_\mrm{p}^{-1} \in \RHinf$ such that \eqref{eqn:performance_analysis} holds. Choosing $N_{K_\mrm{p}} = \tilde{N}_{K_\mrm{p}}\alpha_\mrm{p}$, $D_{K_\mrm{p}} = \tilde{D}_{K_\mrm{p}}\alpha_\mrm{p}$ results in $K_\mrm{p} = N_{K_\mrm{p}}D_{K_\mrm{p}}^{-1} = \tilde{N}_{K_\mrm{p}}\tilde{D}_{K_\mrm{p}}^{-1}$ and consequently \ref{thm:karimi_b} holds.
	
	$(\Leftarrow)$ Assume \ref{thm:karimi_b} holds. Because $D_\mrm{p} \in \RHinf$ and $D_{\mrm{p}}(i\omega)$ is positive for all $\omega \in \Omega$, $D_{\mrm{p}}(i\omega)$ cannot encircle the origin when $\omega$ traverses the Nyquist contour. Thus $D_{\mrm{p}}^{-1} \in \RHinf$ and $K_\mrm{p}$ internally stabilizes $G_\mrm{p}$. Furthermore,
	\begin{equation}
		\begin{split}
			 \lvert D_\mrm{p}(i\omega) \rvert \geq \Re \{ D_{\mrm{p}}(i\omega) \} \\ 
			 \forall \omega \in \Omega, \, \forall \mrm{p} \in \mbb{P}
		\end{split}
	\end{equation}
	implies
	\begin{equation}
		\begin{split}
			\lvert D_\mrm{p}(i\omega) \rvert > \gamma^{-1} \lvert W_T(i\omega)N_{\mrm{p}}(i\omega) \rvert \\ 
			\forall \omega \in \Omega, \, \forall \mrm{p} \in \mbb{P}
		\end{split}
	\end{equation}
	and consequently \ref{thm:karimi_a} holds. This completes the proof.	
\end{document}